\documentclass[10pt, twocolumn, pre, aps, superscriptaddress, showpacs]{revtex4-1}
\usepackage{amsmath, graphicx, subfigure, tikz}
\begin{document}

    \title{Maximally Random Discrete-Spin Systems with Symmetric and Asymmetric Interactions and Maximally Degenerate Ordering}
    \author{Bora Atalay}
    \affiliation{Faculty of Engineering and Natural Sciences, Sabanc\i\ University, Tuzla, Istanbul 34956, Turkey}
    \author{A. Nihat Berker}
    \affiliation{Faculty of Engineering and Natural Sciences, Kadir Has University, Cibali, Istanbul 34083, Turkey}
    \affiliation{Department of Physics, Massachusetts Institute of Technology, Cambridge, Massachusetts 02139, USA}

%05.10.Cc    Renormalization group methods
%64.60.ae    Renormalization-group theory
%64.60.De    Statistical mechanics of model systems
%75.10.Nr    Spin-glass and other random models
%75.50.Lk    Spin glasses and other random magnets

%64.60.Cn    Order-disorder transformations
%05.50.+q    Lattice theory and statistics (Ising, Potts, etc.)
%61.43.-j    Disordered solids
%75.10.Hk    Classical spin models

\begin{abstract}

Discrete-spin systems with maximally random nearest-neighbor
interactions that can be symmetric or asymmetric, ferromagnetic or
antiferromagnetic, including off-diagonal disorder, are studied, for
the number of states $q=3,4$ in $d$ dimensions. We use
renormalization-group theory that is exact for hierarchical lattices
and approximate (Migdal-Kadanoff) for hypercubic lattices. For all
$d>1$ and all non-infinite temperatures, the system eventually
renormalizes to a random single state, thus signaling $q \times q$
degenerate ordering.  Note that this is the maximally degenerate
ordering. For high-temperature initial conditions, the system crosses
over to this highly degenerate ordering only after spending many
renormalization-group iterations near the disordered
(infinite-temperature) fixed point. Thus, a temperature range of
short-range disorder in the presence of long-range order is
identified, as previously seen in underfrustrated Ising spin-glass
systems.  The entropy is calculated for all temperatures, behaves
similarly for ferromagnetic and antiferromagnetic interactions, and
shows a derivative maximum at the short-range disordering
temperature. With a sharp immediate contrast of infinitesimally
higher dimension $1+\epsilon$, the system is as expected disordered
at all temperatures for $d=1$.

\end{abstract}
\maketitle

\section{Introduction: Asymmetric and Symmetric Maximally Random Spin Models}

Spin models such as Ising, Potts, ice models show a richness of
phase transitions and multicritical phenomena \cite{Nienhuis,
Delfino} that is qualitatively compounded with the addition of
frozen (quenched) randomness. Examples are the emerging chaos in
spin glasses with competing ferromagnetic and antiferromagnetic (and
more recently, without recourse to ferromagnetism vs.
antiferromagnetism, competing left and right chiral \cite{Caglar3})
interactions, the conversion of first-order phase transitions to
second-order phase transitions, and the infinite multitude of
accumulating phases as devil's staircases. In the current study,
frozen randomness is taken to the limit, in $q=3,4$ state models in
arbitrary dimension $d$ and the results are quite unexpected.

Thus, changes in the critical properties and the phase-transition
order are the effects quenched randomness, as well as the appearance
of new phenomena such as chaotic rescaling and devil's staircase
topologies of phase diagrams.  A key microscopic ingredient in these
phenomena is the occurrence of frustration, in which all
interactions along closed paths in the lattice cannot be
simultaneously satisfied.  The renormalization-group transformation
that we use in this study is equipped to study frustrated systems
(and thus has been extensively used in spin-glass systems), as can
be seen below by from the equivalent hierarchical lattice where
closed loops occur corresponding to bond-moving following
decimation.

\begin{figure*}[ht!]
\centering
\includegraphics[scale=0.78]{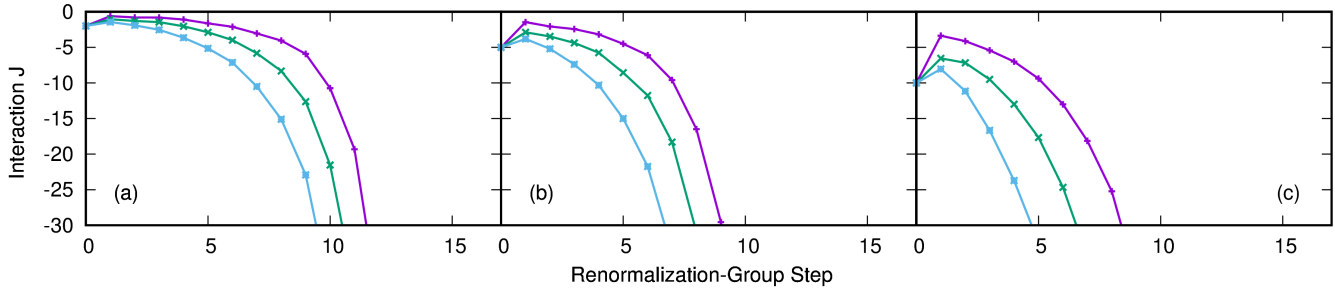}
\caption{Renormalization-group trajectories for the system with $q =
3$ states in $d = 3$ dimensions, starting at three different
temperatures $T = J^{-1}$ from Eqs. (2,3), namely starting with (a)
$J=0.02$, (b) $J=0.20$, (c) $J=0.50$. Shown are the second $(J_2)$,
third $(J_3)$ largest values and the matrix average of the eight
non-leading energies $<J_{2-9}>$ of the transfer matrix (Eq. (4)),
averaged over the quenched random distribution. The leading energy
is $J_1 = 0$ by subtractive choice. The different starting values
can be seen on the left axis of each panel (corresponding to
renormalization-group step 0). Starting at any non-zero temperature,
the system renormalizes to a state in which the leading energy is
totally dominant, all other energies renormalizing to $-\infty$. The
matrix position of the single asymptotically dominant element occurs
randomly among the $q \times q$ possibilities including off-diagonal
and therefore necessarily asymmetric, but is the same across the
quenched random distribution. However, starting at high
temperatures, as seen e.g. in the left and center panels in this
figure, the system spends many renormalization-group iterations near
the infinite-temperature fixed point (where all energies are zero),
before crossing over to the ordered fixed point.  Since the energies
at a specific step of a renormalization-group trajectory directly
show the effective couplings across the length scale that is reached
at that renormalization-group step, this behavior indicates islands
of short-range disorder at the short length scales that correspond
to the initial steps of a renormalization-group trajectory.  These
islands of short-range disorder nested in long-range order have been
explicitly calculated and shown in spin-glass systems in Ref. (20).
These islands of short-range disorder occur in the presence of
long-range order, since the trajectories eventually flow to the
strong-coupling fixed point.  As temperature is increased (changing
the renormalization-group initial condition), these short-range
disordered regions order, giving rise to the smooth specific heat
peak, but no phase transition singularity, as there is no additional
fixed-point structure underlying this short-range ordering.}
\end{figure*}

The systems that we study are quenched maximally random $q$-state
discrete spin models with nearest-neighbor interactions, with
Hamiltonian
\begin{equation}
\centering - \beta {\cal H} = -\sum_{\left<ij\right>} \beta {\cal
H}_{ij},
\end{equation}
where the sum is over nearest-neighbor pairs of sites $<ij>$.

The maximal randomness is best expressed in the transfer matrix
$T_{ij}$, e.g., for $q=3$,
\begin{multline}
\textbf{T}_{ij} \equiv e^{-\beta {\cal H}_{ij}} =\\
\left(
\begin{array}{ccc}
1 & e^J & 1 \\
1 & 1 & e^J \\
e^J & 1 & 1 \end{array} \right), \left(
\begin{array}{ccc}
1 & 1 & e^J \\
e^J & 1 & 1 \\
1 & e^J & 1 \end{array} \right), \left(
\begin{array}{ccc}
e^J & 1 & 1 \\
1 & 1 & e^J \\
1 & e^J & 1 \end{array} \right),\\
\left(
\begin{array}{ccc}
1 & 1 & e^J \\
1 & e^J & 1 \\
e^J & 1 & 1 \end{array} \right),\left(
\begin{array}{ccc}
1 & e^J & 1 \\
e^J & 1 & 1 \\
1 & 1 & e^J \end{array} \right), \, \text{or} \left(
\begin{array}{ccc}
e^J & 1 & 1 \\
1 & e^J & 1 \\
1 & 1 & e^J \end{array} \right),
\end{multline}
where each row and each column has, randomly, a single $e^J$
element, so that there are 6 such possibilities (for $q=4$, also
studied here, there are 24 such possibilities), and $J>0$ or $J<0$
respectively for ferromagnetic or antiferromagnetic interactions,
both of which are treated in this study.  The last matrix
corresponds to the usual Potts model.  In fact, taken by itself as a
pure (non-random) model, each of these transfer matrices can be
mapped to a Potts model by relabeling the spin states in one of the
two sublattices, in hypercubic lattices and corresponding
hierarchical lattices.  Thus, for the ferromagnetic case, for $d>1$,
a low-temperature ferromagnetic phase and a high-temperature
disordered phase occurs.  For the antiferromagnetic case, the
low-temperature phase is a critical phase and appears at a higher
dimension.\cite{BerkerKadanoff1,BerkerKadanoff2}

In Hamiltonian terms, the currently studied, quenched random model
is
\begin{equation}
-\beta{\cal H}_{i,j}=J\delta_{\sigma_i,P(\sigma_j)},
\end{equation}
where $P$ is a random permutation of $\{a,b,c\}$. Thus, at a given
site $i$, for a given spin state, say $s_i = a$, randomly any one of
the spin states $s_j = a,b,$ or $c$ of the nearest-neighbor site $j$
is energetically favored (unfavored) for ferromagnetic
(antiferromagnetic) interactions.  This favor (unfavor) is
independently random for each of the nearest-neighbors $j$.  Under
renormalization-group transformation, all elements of the transfer
matrices across the system randomize.  Therefore, we have not
included in our renormalization-group initial conditions the cases
where there is a difference between the less favored two states, to
keep the enunciation of the model simple. However, since our
renormalization-group trajectories traverse the latter states, we
are confident that our results will not be affected by such a
sub-discrimination.

The first two possible transfer matrices on the right side of Eq.
(1) represent asymmetric interaction, in the sense that the
nearest-neighbor states $(s_i,s_j)=(a,b)$ and $(b,a)$ have different
energies, where $s_i=a,b,$ or $c$ are the $q=3$ possible states of a
given site $i$. Asymmetric interactions occur in neural network
systems \cite{Domany} and are largely unexplored in statistical
mechanics. On the other hand, the last four possible transfer
matrices on the right side of Eq. (1) exemplify symmetric
interaction, the nearest-neighbor states $(s_i,s_j)=(a,b)$ and
$(b,a)$ having the same energies.  As also explained below, even
when starting with only symmetric interactions (the last four
matrices), asymmetric interactions are generated under
renormalization-group transformations (as can be seen, e.g. by
multiplying the third and fifth matrices in Eq. (2), corresponding
to a renormalization-group decimation) and the same ordering results
are obtained.  Thus, asymmetric interactions are generated by
off-diagonal (symmetric) disorder.  The generalization of the above
model to arbitrary $q$ is obvious.

\section{Renormalization-Group Transformation}

The renormalization-group method is readily implemented to the
transfer matrix form of the interactions.  The quenched randomness
aspect of the problem is included by randomly creating 500 transfer
matrices from the 6 possibilities of Eq. (1) and perpetuating these
random 500 transfer matrices throughout the renormalization-group
steps given below.  Note that we start with a single initial value
of $J$, which is proportional to the inverse temperature.  Quenched
randomness comes from the positioning within the matrix.  Under
renormalization-group transformation, each matrix element evolves
quantitively quenched randomly.

\begin{figure}[ht!]
\centering
\includegraphics[scale=0.95]{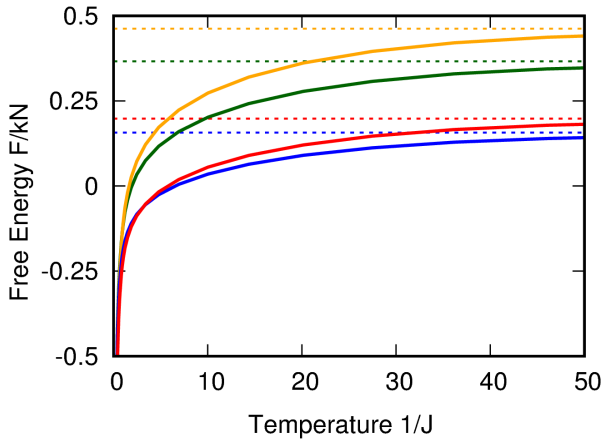}
\caption{Calculated free energy per bond as a function of
temperature $T=J^{-1}$. The curves are, from top to bottom, for
$(q=4,d=2),(q=3,d=2),(q=4,d=3),(q=3,d=3)$. The expected $T=\infty$
values of $ f = F / kN = \ln q / (b^d-1)$ are given by the dashed
lines and match the calculations.}
\end{figure}

The renormalization-group transformation begins by the "bond-moving"
step in which $b^{d-1}$ transfer matrices, each randomly chosen from
the 500, have their corresponding matrix elements multiplied.  This
operation is repeated 500 times, thus generating 500 new transfer
matrices.  The final, "decimation" step of the renormalization-group
transformation is the matrix multiplication of $b$ transfer
matrices, again each randomly chosen from the 500.  This operation
is also repeated 500 times, again generating 500 renormalized
transfer matrices.  The length rescaling factor is taken as $b=2$ in
our calculation.  At each transfer-matrix calculation above, each
element of the resulting transfer matrix is divided by the largest
element, resulting in a matrix with the largest element being unity.
This does not affect the physics, since it corresponds to
subtracting a constant from the Hamiltonian.  These subtractive
constants (the natural logarithm of the dividing element) are
scale-accumulated, as explained below, for the calculation of
entropy.

The above transformation is the approximate Migdal-Kadanoff
\cite{Migdal,Kadanoff} renormalization-group transformation for
hypercubic lattices and, simultaneously, the exact
renormalization-group transformation of a hierarchical lattice
\cite{BerkerOstlund,Kaufman1,Kaufman2}. This procedure has been
explained in detail in previous works.\cite{Caglar3} For most recent
exact calculations on hierarchical lattices, see Ref.\cite{Masuda,
Boettcher4, Bleher, Zhang2017, Peng, Nogawa, Sirca, Maji}, including
finance \cite{Sirca} and DNA-binding \cite{Maji} problems.

\begin{figure}[ht!]
\centering
\includegraphics[scale=0.95]{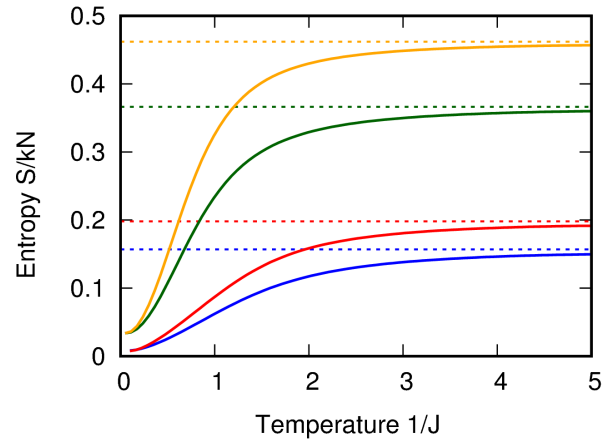}
\caption{Calculated entropy per bond as a function of temperature
$T=J^{-1}$, for $q=3,4$ states in $d=3,4$ dimensions. The curves
are, from top to bottom, for
$(q=4,d=2),(q=3,d=2),(q=4,d=3),(q=3,d=3)$. The expected $T=\infty$
values of $ S/kN = \ln q / (b^d-1)$ are given by the dashed lines
and match the calculations.}
\end{figure}

\section{Asymptotically Dominant All-Temperature Freezing in $d>1$ with High-Temperature Short-Range Disordering}

Figure 1 shows the renormalization-group trajectories for the system
with $q=3$ states in $d=3$ dimensions, starting at three different
temperatures $T=J^{-1}$, where $J$ refers to the
renormalization-group-trajectory initial conditions shown in
Eqs.(2,3). Shown are the second ($J_2$) and third ($J_3$) largest
values of the energies (dimensionless, being temperature-divided)
that appear exponentiated in the transfer matrix elements,
\begin{equation}
J_{ij} = \ln (T_{ij}),
\end{equation}
averaged over the quenched random distribution, where $T_{ij}$ are
the elements of the $q \times q$ transfer matrix $\textbf{T}_{ij}$.
The matrix average of the eight non-leading energies $<J_{2-9}>$,
averaged over the quenched random distribution, is also shown.  The
leading energy is $J_1=0$ by subtractive overall constant, as
explained above. As seen in this figure, starting at low temperature
$T=2$, the system renormalizes to a state in which the leading
energy is totally dominant, all other energies renormalizing to
$-\infty$. The matrix position of the single asymptotically dominant
element occurs randomly among the $q \times q$ possibilities
including off-diagonal and therefore necessarily asymmetric, but is
the same across the quenched random distribution. The number of
possible dominant transfer-matrix elements gives the degeneracy of
the ordered phase, so that with $q \times q$, maximal degeneracy is
achieved.  A diagonal element of the transfer matrix being dominant
means that one state, e.g. $s_i = c$ dominates at the strong
coupling fixed point and characterizes the ordered phase. This does
have the usual permutational symmetry of the Potts model, being
physically equivalent to all diagonal elements dominating, but with
non-diagonal elements zero so that only one spin state dominates the
entire physical system.  The equivalence is not complete only in the
fact that the latter picture allows different domains in the system,
where as the former does not.  A non-diagonal element $T_{km} = 1$
being dominant maintains itself by having $T_{im}, T_{kj}$, where
$i\neq k, j\neq m$, being small, decreasing under
renormalization-group, but non-zero.  The corresponding spin state
is highly degenerate, as can be seen from the renormalization-group
solution, where each spin has a degeneracy of 2 (still less than the
disordered number of $q$), seen at decimation transformations, and
the system is randomly populated by two spin states corresponding to
the indices $k$ and $m$ of the dominant $T_{km}$.

\begin{figure}[ht!]
\centering
\includegraphics[scale=0.95]{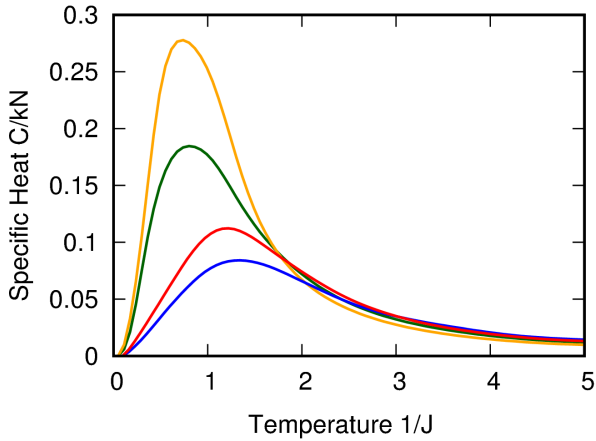}
\caption{Calculated specific heat as a function of temperature
$T=J^{-1}$, for $q=3,4$ states in $d=3,4$ dimensions. The curves
are, from top to bottom, for
$(q=4,d=2),(q=3,d=2),(q=4,d=3),(q=3,d=3)$. A specific heat maximum
occurs at short-range disordering.}
\end{figure}

\begin{figure}[ht!]
\centering
\includegraphics[scale=0.95]{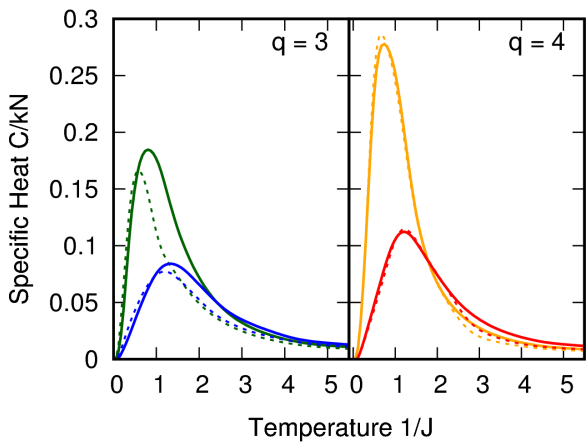}
\caption{Calculated specific heat as a function of temperature
$T=|J|^{-1}$ for ferromagnetic $(J>0)$, full curves, and
antiferromagnetic $(J<0)$, dashed curves, systems, for $q=3,4$
states in $d=3,4$ dimensions. The curves are, from top to bottom in
each panel, for $d=2$ and $d=3$. The quantitatively same short-range
disordering behavior is seen for both ferromagnetic and
antiferromagnetic systems.}
\end{figure}

Moreover, starting at high temperatures, as seen e.g. in the left
and center panels of Fig. 1, the system spends many
renormalization-group iterations near the infinite-temperature fixed
point (where all energies are zero), before crossing over to the
ordered fixed point. This signifies short-range disorder, in the
presence of long-range order, as also reflected in the specific heat
peaks caused by short-range disordering as discussed below. A
similar smeared transition to short-range disorder in the presence
of long-range order has previously been seen in underfrustrated
Ising spin-glass systems.\cite{Ilker2}

We have repeated our calculations for non-integer spatial dimensions
approaching $d=1$ from above, by keeping the bond-moving number
$b^{d-1} = 2$ and increasing the decimation number $b$.  The
behavior described above obtains for all $d\gtrsim 1$, albeit with
an increasing high-temperature range of short-range disorder, and
higher number of renormalization-group steps to strong coupling, as
$d=1$ is approached.  At $d=1$, the infinite-temperature fixed point
is the sole attractor and the system is disordered at all
temperatures.

\section{Free Energy, Entropy, and Specific Heat}

The renormalization-group solution gives the complete equilibrium
thermodynamics for the systems studied.  The dimensionless free
energy per bond $f = F/kN$ is obtained by summing the additive
constants generated at each renormalization-group step,
\begin{equation}
f \, = \, \frac{1}{N} \ln \sum_{\{s_i\}} e^{-\beta {\cal H}} \, = \,
\sum_{n=1} \frac{G^{(n)}}{b^{dn}},
\end{equation}
where $N$ is the number of bonds in the initial unrenormalized
system, the first sum is over all states of the system, the second
sum is over all renormalization-group steps $n$, $G^{(n)}$ is the
additive constant generated at the $(n)$th renormalization-group
transformation averaged over the quenched random distribution, and
the sum quickly converges numerically.

From the dimensionless free energy per bond $f$, the entropy per
bond $S/k N$  is calculated as
\begin{equation}
\frac{S}{k N} = f -J \frac{\partial f}{\partial J}
\end{equation}
and the specific heat $C/k N$ is calculated as
\begin{equation}
\frac{C}{k N} = T \frac{\partial (S/k N)}{\partial T} = - J
\frac{\partial (S/k N)}{\partial J}\,.
\end{equation}
Figures 2-4 give the calculated free energies $f$, entropies $S/kN$,
and specific heats $C/kN$ per bond as functions of temperature
$T=J^{-1}$, for $q=3,4$ states in $d=3,4$ dimensions. The expected
$T=\infty$ values of $ f = \ln q / (b^d-1)$ and $ S/kN = \ln q /
(b^d-1)$ are given by the dashed lines and match the calculations.

As explained in Fig. 1, the specific heat maximum occurs at the
temperature of the short-range disordering.  In this figure,
starting at high temperatures, as seen e.g. in the left and center
panels in this figure, the system spends many renormalization-group
iterations near the infinite-temperature fixed point (where all
energies are zero), before crossing over to the ordered fixed point.
Since the energies at a specific step of a renormalization-group
trajectory directly show the effective couplings across the length
scale that is reached at that renormalization-group step, this
behavior indicates islands of short-range disorder at the short
length scales that correspond to the initial steps of a
renormalization-group trajectory.  These islands of short-range
disorder nested in long-range order have been explicitly calculated
and shown in spin-glass systems in Ref. (20). These islands of
short-range disorder occur in the presence of long-range order,
since the trajectories eventually flow to the strong-coupling fixed
point.  As temperature is increased (changing the
renormalization-group initial condition), these short-range
disordered regions order, giving rise to the smooth specific heat
peak, but no phase transition singularity, as there is no additional
fixed-point structure underlying this short-range ordering. Specific
heat maxima away from phase transitions, due to short-range
ordering, have been calculated in a variety of
systems.\cite{BerkerNelson,Ilker2}

\section{Antiferromagnetic Maximally Random Systems}

We have repeated our calculations for antiferromagnetic $(J<0)$
systems and obtained quantitatively similar behavior. Figs. 5 shows
the calculated specific heats as a function of temperature
$T=|J|^{-1}$ for ferromagnetic $(J>0)$ and antiferromagnetic $(J<0)$
systems, for $q=3,4$ states in $d=3,4$ dimensions. The
full-temperature range $(T < -\infty)$ maximally degenerate
long-range ordering and a quantitatively same short-range
disordering at high temperature is seen for both ferromagnetic and
antiferromagnetic systems.

\section{Conclusion}
We have studied maximally random discrete-spin systems with
symmetric and asymmetric interactions and have found, quite
surprisingly, (1) quenched random long-range order at all
non-infinite temperatures for $d > 1$, (2) short-range disordering
at high temperatures, via a smeared transition and a specific-heat
peak, while sustaining long-range order.  The latter behavior has
also been seen in underfrustrated Ising spin-glass
systems.\cite{Ilker2}

\begin{acknowledgments}
Support by the Academy of Sciences of Turkey (T\"UBA) is gratefully
acknowledged. We thank Tolga \c{C}a\u{g}lar for most useful
discussions.
\end{acknowledgments}


\begin{references}

\bibitem{Nienhuis} B. Nienhuis, A. N. Berker, E. K. Riedel, and M. Schick, Phys. Rev. Let. {\bf 43}, 737 (1979).
\bibitem{Delfino} G. Delfino and E. Tartaglia, Phys. Rev. E {\bf 96}, 042137 (2017).
\bibitem{Caglar3} T. \c{C}a\u{g}lar and A. N. Berker, Phys. Rev. E {\bf 96}, 032103 (2017).
\bibitem{BerkerKadanoff1} A. N. Berker and L. P. Kadanoff, J. Phys. A {\bf 13}, L259 (1980).
\bibitem{BerkerKadanoff2} A. N. Berker and L. P. Kadanoff, J. Phys. A {\bf 13}, 3786 (1980).
\bibitem{Domany} E. Domany, J. L. van Hemmen, and K. Schulten (Eds.) \textit{Models of Neural Networks}
(Springer-Verlag, Berlin, 1991).

\bibitem{Migdal} A. A. Migdal, Zh. Eksp. Teor. Fiz. {\bf69}, 1457 (1975) [Sov. Phys. JETP {\bf42}, 743 (1976)].
\bibitem{Kadanoff} L. P. Kadanoff, Ann. Phys. (N.Y.) {\bf100}, 359 (1976).

\bibitem{BerkerOstlund} A. N. Berker and S. Ostlund, J. Phys. C {\bf 12}, 4961 (1979).
\bibitem{Kaufman1} R. B. Griffiths and M. Kaufman, Phys. Rev. B {\bf 26}, 5022R (1982).
\bibitem{Kaufman2} M. Kaufman and R. B. Griffiths, Phys. Rev. B {\bf 30}, 244 (1984).

\bibitem{Masuda} N. Masuda, M. A. Porter, and R. Lambiotte, Phys. Repts. {\bf 716}, 1 (2017).
\bibitem{Boettcher4} S. Li and S. Boettcher, Phys. Rev. A {\bf 95}, 032301 (2017).
\bibitem{Bleher} P. Bleher, M. Lyubich, and R. Roeder, J. Mathematiques Pures et Appliqu\'{e}es {\bf 107}, 491 (2017).
\bibitem{Zhang2017} H. Li and Z. Zhang, Theoretical Comp. Sci. {\bf 675}, 64 (2017).
\bibitem{Peng} J. Peng and E. Agliari, Chaos {\bf 27} 083108 (2017).
\bibitem{Nogawa} T. Nogawa, arXiv:1710.04014 [cond-mat.stat-mech](2017)

\bibitem{Sirca} S. J. Sirca and M. Omladic, ARS Mathematica Contemporanea {\bf 13}, 63 (2017).
\bibitem{Maji} J. Maji, F. Seno, A. Trovato, and S. M. Bhattacharjee, J. Stat. Mech.: Theory Exp. 073203 (2017).

\bibitem{Yesilleten} D. Ye\c{s}illeten and A. N. Berker, Phys. Rev. Lett. Phys. Rev. Lett. {\bf 78}, 1564 (1997).
\bibitem{BerkerNelson} A. N. Berker and D. R. Nelson, Phys. Rev. B {\bf 19}, 2488 (1979).
\bibitem{Ilker2} E. Ilker and A. N. Berker, Phys. Rev. E {\bf 89}, 042139 (2014).

\end{references}
\end{document}